# Applicability of Partial Ternary Full Adder in Ternary Arithmetic Units

Aida Ghorbani Asibelagh [1], Reza Faghih Mirzaee [2*]

[1] Department of Computer Engineering, North Tehran Branch, Islamic Azad University, Tehran, Iran

[2] Department of Computer Engineering, Shahr-e-Qods Branch, Islamic Azad University, Tehran, Iran

* Corresponding author's e-mail: r.f.mirzaee@qodsiau.ac.ir; Tel: +98-21-46896000 (1261)

*Abstract:* This paper explores whether or not a complete ternary full adder, whose input variables can independently be '0', '1', or '2', is indispensable in the arithmetic blocks of adder, subtractor, and multiplier. Our investigations show that none of the mentioned arithmetic units require a complete ternary full adder. Instead, they can be designed by use of partial ternary full adder, whose input carry never becomes '2'. Furthermore, some new ternary compressors are proposed in this paper without the requirement of complete ternary full adder. The usage of partial ternary full adder can help circuit designers to simplify their designs, especially in transistor level.

# I. Introduction

Full adder is undoubtedly the most important block in computer arithmetic. It is the core element of other arithmetic operations. This is the reason why it has received so much attention in the literature. It is almost always situated along the critical path of a system, affecting its clock frequency. Therefore, its improvement leads to a major influence on the performance of the whole system [1, 2]. It takes three equally-weighted input variables and computes their summation. Two outputs are generated as a result of a single-bit addition:

1. A *Sum*, with the same bit position of the inputs.

2. An *Output Carry* ($C_{out}$), with one higher bit position than *Sum*.

Full adders are put next to each other to be able to add two numbers. Ripple carry adder (RCA) is the most well-known structure for this purpose [3]. Full adder is also widely used in compressors and multipliers [3, 4]. Subtraction is another basic mathematical operation which is often converted to addition. Finally, division algorithms are usually based on either subtraction or multiplication iterative operators [5]. Therefore, full adder plays an important role in the four basic mathematical operations.

Theoretically, ternary computations can be done a factor of $\log_2(3)$ (≈1.585) faster than binary since a ternary digit (Trit) is equal to $\log_2(3)$ bits of information [6]. Unbalanced ternary logic is a numeral system with the number set of $\{0, 1, 2\}_3$. It is known as the extended binary notation [6, 7]. An ordinary ternary full adder (TFA) adds three ternary digits, each of which can independently be '0', '1', or '2'. Such a regular TFA is called a *complete* one in this paper. However, as indicated in [8] and also fully discussed in [9], the input carry ($C_{in}$) never becomes '2' in a ternary RCA. Thus, some unnecessary transistors, which make a route when the input carry is '2', can be eliminated. Their elimination leads to a *partial* TFA with smaller area and lower power consumption.

This paper explores whether it is applicable to use partial TFAs in some arithmetic blocks such as subtractor and multiplier. If so, TFAs can be designed with fewer transistors. In addition, some new ternary compressors are presented without the requirement of a complete TFA. The rest of the paper is organized as follows: Section 2 reviews ternary logic briefly. The necessity of a complete TFA in adder, subtractor, multiplier, and compressor is studied in Section 3. Finally, Section 4 concludes the paper.

# II. Ternary Logic in Brief

Multiple-valued logic (MVL) is a propositional calculus in which there are more than two logic values. Amongst different MVL systems, ternary logic (or three-valued logic) is the most efficient one, providing lower cost and complexity than binary [10, 11]. There are two types of ternary systems:

1. Balanced (or signed) ternary, with the number set of $\{-1, 0, 1\}_3$.

2. Unbalanced (or unsigned) ternary, with the number set of $\{0, 1, 2\}_3$.

Both numeral systems are irredundant. It means that a number has a unique representation. Table 1 illustrates the representation of decimal digits in both ternary systems. There are three main differences between them:

1. Due to the existence of the negative digit −1 (or $\bar{1}$), there is no need of a minus sign to represent a negative number in the balanced ternary logic.

2. As exemplified in Table 1, in most cases, a number in the unbalanced ternary logic can be written with fewer digits.

3. A negative voltage source is required to implement the balanced ternary logic in digital electronics.

**Table I:** Representation of Decimal Digits in Ternary Number Systems

| Decimal | Balanced Ternary | Unbalanced Ternary |
|---|---|---|
| 0 | 0 | 0 |
| 1 | 1 | 1 |
| 2 | 1$\bar{1}$ | 2 |
| 3 | 10 | 10 |
| 4 | 11 | 11 |
| 5 | 1$\bar{1}\bar{1}$ | 12 |
| 6 | 1$\bar{1}$0 | 20 |
| 7 | 1$\bar{1}$1 | 21 |
| 8 | 10$\bar{1}$ | 22 |
| 9 | 100 | 100 |

The unsigned ternary logic has gained much more attention in the literature mainly because of the last two above reasons. Ternary logic attempts to reduce the amount of interconnections inside and outside a chip [12]. Inclusion of an extra negative voltage source is in contrast with the primary goal of ternary logic. It imposes additional wiring and routing complexities in VLSI design. Moreover, the unbalanced version is more human-readable representation.

Logical conjunction (AND), disjunction (OR), and negation (NOT) can be calculated in ternary logic by Eqs. (1) to (3), respectively.

$$AND(a,b) = \min(a,b) \qquad (1)$$

$$OR(a,b) = \max(a,b) \qquad (2)$$

$$NOT(a) = 2 - a \qquad (3)$$

# III. Partial Ternary Full Adder in Arithmetic Circuits

## 3.1. Addition (Review)

A ternary RCA (Fig. 1a) is made up of a ternary half adder (THA) and a sequence of TFAs, whose truth tables are depicted in Tables 2 and 3, respectively. It adds two ternary numbers such as $A:a_{n-1}...a_2a_1a_0$ and $B:b_{n-1}...b_2b_1b_0$. The first component in RCA is a THA since no input carry enters the least significant column. As it is shown in Table 2, the output carry of a THA belongs to the set $\{0, 1\}_3$. Consequently, the input carry received at the first TFA never becomes '2'. In this situation, the first complete TFA can be replaced with a partial one, whose truth table is demonstrated in Table 4. Subsequently, the output carry of the first partial TFA does not produce '2' either. As a result, the second as well as the rest of the complete TFAs can be replaced with partial ones as well (Fig. 1b) [8, 9].

**Table II:** Truth Table of THA

| a | b | $C_{out}$ | Sum |
|---|---|---|---|
| 0 | 0 | 0 | 0 |
| 0 | 1 | 0 | 1 |
| 0 | 2 | 0 | 2 |
| 1 | 0 | 0 | 1 |
| 1 | 1 | 0 | 2 |
| 1 | 2 | 1 | 0 |
| 2 | 0 | 0 | 2 |
| 2 | 1 | 1 | 0 |
| 2 | 2 | 1 | 1 |

**Table III:** Truth Table of Complete TFA

| a | b | $C_{in} = 0$ | | $C_{in} = 1$ | | $C_{in} = 2$ | |
|---|---|---|---|---|---|---|---|
| | | $C_{out}$ | Sum | $C_{out}$ | Sum | $C_{out}$ | Sum |
| 0 | 0 | 0 | 0 | 0 | 1 | 0 | 2 |
| 0 | 1 | 0 | 1 | 0 | 2 | 1 | 0 |
| 0 | 2 | 0 | 2 | 1 | 0 | 1 | 1 |
| 1 | 0 | 0 | 1 | 0 | 2 | 1 | 0 |
| 1 | 1 | 0 | 2 | 1 | 0 | 1 | 1 |
| 1 | 2 | 1 | 0 | 1 | 1 | 1 | 2 |
| 2 | 0 | 0 | 2 | 1 | 0 | 1 | 1 |
| 2 | 1 | 1 | 0 | 1 | 1 | 1 | 2 |
| 2 | 2 | 1 | 1 | 1 | 2 | 2 | 0 |

**Table IV:** Truth Table of Partial TFA

| a | b | $C_{in}=0$ | | $C_{in}=1$ | | $C_{in}=2$ | |
|---|---|---|---|---|---|---|---|
| | | $C_{out}$ | Sum | $C_{out}$ | Sum | $C_{out}$ | Sum |
| 0 | 0 | 0 | 0 | 0 | 1 | 0 | 2 |
| 0 | 1 | 0 | 1 | 0 | 2 | 1 | 0 |
| 0 | 2 | 0 | 2 | 1 | 0 | 1 | 1 |
| 1 | 0 | 0 | 1 | 0 | 2 | 1 | 0 |
| 1 | 1 | 0 | 2 | 1 | 0 | 1 | 1 |
| 1 | 2 | 1 | 0 | 1 | 1 | 1 | 2 |
| 2 | 0 | 0 | 2 | 1 | 0 | 1 | 1 |
| 2 | 1 | 1 | 0 | 1 | 1 | 1 | 2 |
| 2 | 2 | 1 | 1 | 1 | 2 | 2 | 0 |

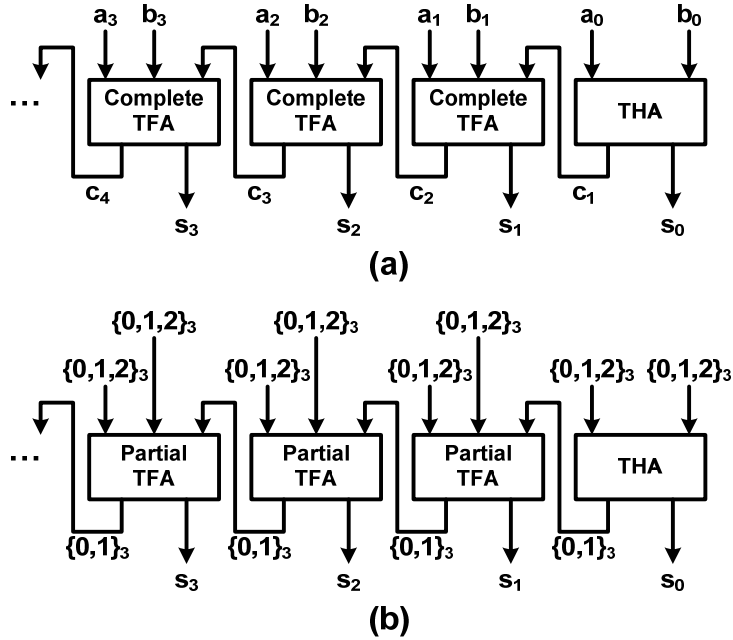

**Figure 1.** Ternary adder made by, (a) complete TFAs, (b) partial TFAs [9]

*3.2. Subtraction*

A ternary RCA can entirely be made up of partial TFAs since a THA is situated in the least significant column. What if we are obliged to insert a TFA in the first column, e.g. in a subtractor?

Although it is feasible to use ternary half and full subtractors (THS and TFS) for the construction of a ripple borrow subtractor (RBS) [13], subtraction is often converted to addition by means of Eq. (4). This is the same method utilized in [14] to design a ternary adder/subtractor. A ternary subtractor is shown in Fig. 2. It is only composed of partial TFAs. Since the input carry within the first TFA is always '1', the output carry never becomes '2'. Therefore, complete TFAs are not required to perform ternary subtraction.

$$A - B = A + (-B) = A + \overline{B} + 1 \qquad (4)$$

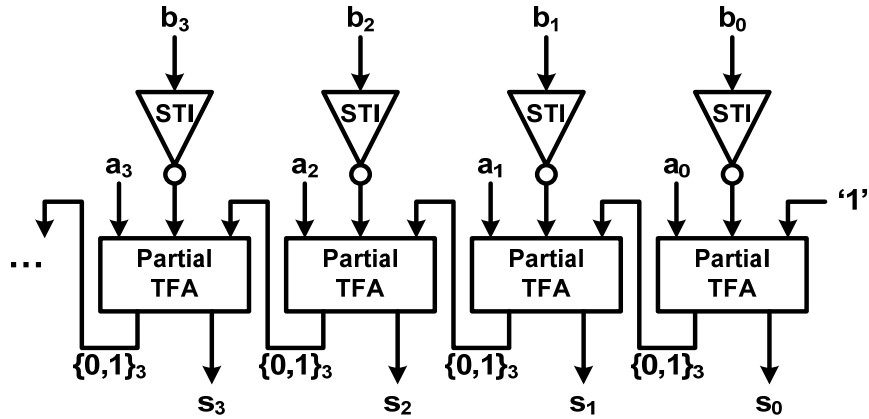

**Figure 2.** Ternary subtractor made by partial TFAs

*3.3. Multiplication*

The truth table of a single-digit ternary multiplier is shown in Table 5. Unlike binary logic, two outputs, *Product* and *Carry*, are generated when two ternary digits are multiplied. The reason is that when 2 is multiplied by another 2, the result cannot be represented by a single trit. The first output might be '0', '1', or '2' (*Product* $\in$ *{0, 1, 2}$_3$*). However, the second one is either '0', or '1' (*Carry* $\in$ *{0, 1}$_3$*). When two ternary numbers are multiplied, several *products* and *carries* are generated in different columns. They are generally called *partial products* (PPs), which have to be summed up and accumulated in the second step of multiplication.

A 6-by-6 digit ternary multiplier is exhibited in Fig. 3. There are three steps in column compression multipliers [15]. At first, PPs are generated. They are divided into *Products*, denoted by dots (•), and *Carries*, denoted by asterisks (*). Then, according to the algorithm by Wallace [16], successive THAs and TFAs accumulate them to the point that only one or two PPs remain within every column (2$^{nd}$ step). The *Sums* and *Carries* created by THAs and TFAs are also marked with dots and asterisks, respectively. Eventually in the 3$^{rd}$ step, a ternary carry-propagate adder such as RCA computes the final result of the multiplication.

Whenever a *Product* is produced in a column during the generation of ternary PPs, a parallel *Carry* is also generated in the adjacent column. As it is shown in Fig. 3, every single column has some dots and some asterisks. This combination offers a chance to avoid the usage of complete TFAs. Since all of the full adders receive at least one asterisk (Fig. 3), partial TFAs are adequate to perform PP accumulation. THAs and partial TFAs reproduce carries again. Therefore, the blend of dots and asterisks continues to exist until the end. Figure 3 provides an example to show that a ternary multiplier does not require complete TFAs. This is true regardless of the size of the multiplier.

**Table V:** Truth Table of Ternary Multiplier

| a | b | C_out | Sum |
|---|---|-------|-----|
| 0 | 0 | 0 | 0 |
| 0 | 1 | 0 | 1 |
| 0 | 2 | 0 | 2 |
| 1 | 0 | 0 | 1 |
| 1 | 1 | 0 | 2 |
| 1 | 2 | 1 | 0 |
| 2 | 0 | 0 | 2 |
| 2 | 1 | 1 | 0 |
| 2 | 2 | 1 | 1 |

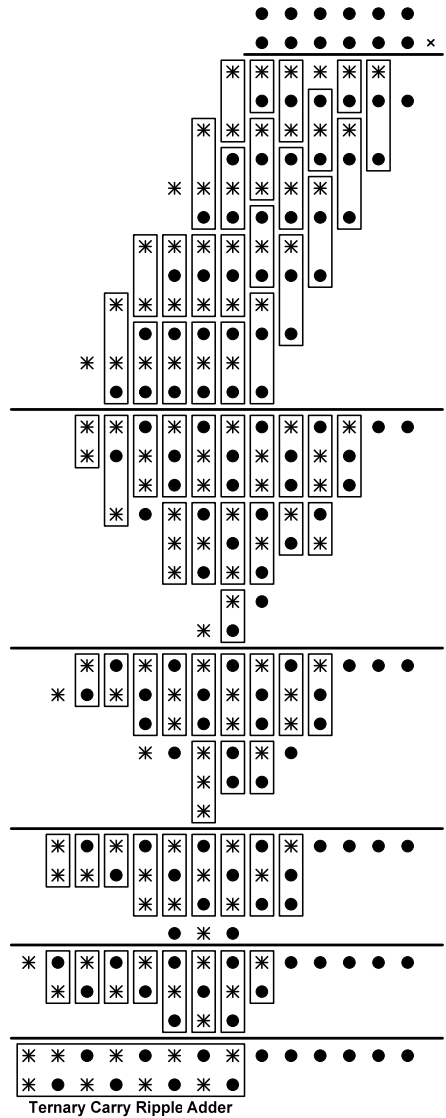

**Figure 3.** Ternary multiplier made by THAs and partial TFAs

*3.4. Compressor*

Compressors can be used for faster accumulation of PPs in the 2nd step of multiplication [17]. In binary logic, wide compressors are usually made up of half and full adders [3, 4]. However, they are not the only main building blocks for the design of high-order compressors in ternary logic. A binary *4:2* compressor is traditionally composed of two full adders (Fig. 4a). On the contrary, a ternary *4:2* compressor (Fig. 4b) is an individual circuit [18]. It loses its efficiency and outlives its usefulness if it is built by two full adders. The reason is that with two outputs in the positions of $3^n$ and $3^{n+1}$ (Fig. 4b), up to four ternary inputs can simply be added ($3^n+3^{n+1}=4\times3^n$). Therefore, there is no need of any input/output carries. However, in the conventional construction of *4:2* compressor (Fig. 4a), the creation of the third output is inevitable.

Compressors with three outputs can accept up to 13 PPs ($3^n+3^{n+1}+3^{n+2}=13\times3^n$). The main target of this paper is to study different situations and scenarios where a complete TFA might be required. In binary logic, full adder plays an important role in constructing high-order compressors. However, complete TFAs are no longer practical for the construction of ternary compressors. Ternary *m:3* ($5\leq m\leq 13$) compressors are proposed in this paper in Fig. 5. They are composed of only ternary *4:2* compressors, THAs, and partial TFAs. In brief:

- *5:3/7:3/11:3/13:3* compressors (Figs. 5a/5c/5g/5i): They do not have any full adder at all. Thus, it does not matter whether the input PPs are denoted by dots or asterisks.

- *6:3/10:3/12:3* compressors (Figs. 5b/5f/5h): The TFA inside these compressors receives a PP, denoted by *, which is either '0' or '1'. Therefore, partial TFAs can be used. There are several such PPs in the second step of a ternary multiplier (Fig. 3).

- *8:3* compressor (Fig. 5d): The TFA receives an output carry signal from a THA, whose output carry never becomes '2' (Table 2). Thus, a partial TFA is used.

- *9:3* compressors (Fig. 5e): The first TFA is partial because it receives at least one PP denoted by *. Consequently, the output carry does not produce '2' (Table 4). Hence, the second TFA can be partial as well.

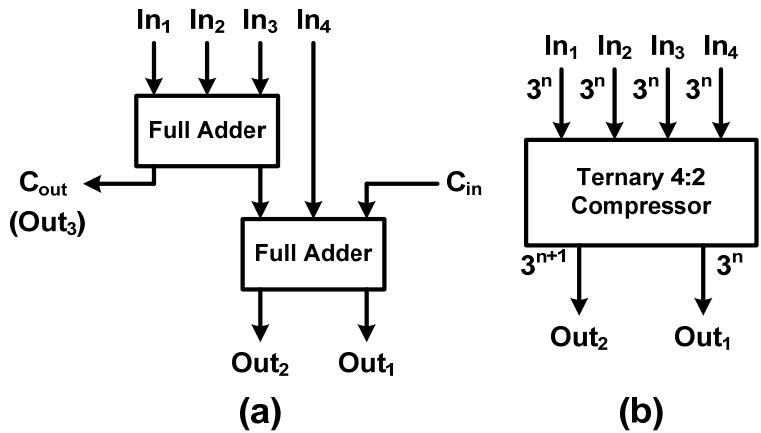

**Figure 4.** 4:2 compressor, (a) Conventional implementation, (b) Ternary 4:2 compressor [18]

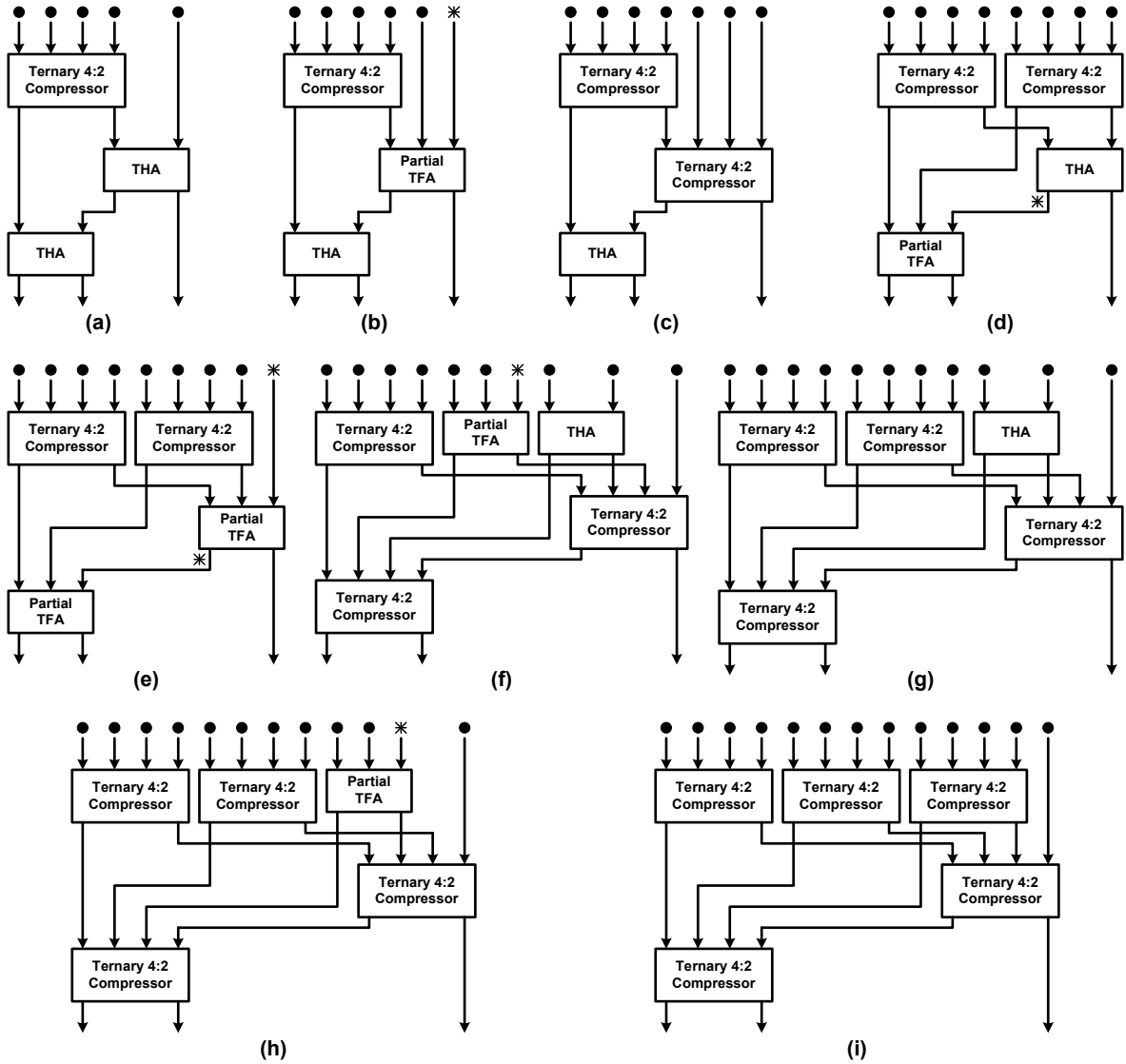

**Figure 5.** Proposed ternary m:3 (5≤m≤13) compressors made by ternary 4:2 compressors, THAs, and partial TFAs, (a) 5:3, (b) 6:3, (c) 7:3, (d) 8:3, (e) 9:3, (f) 10:3, (g) 11:3, (h) 12:3, (i) 13:3

It is worth it again to emphasize the point that, dissimilar to binary logic, TFA is not the principle building block for the construction of wide ternary compressors. The reason is that a complete TFA wastes its output range. The range covered by the outputs of a TFA, *Sum* and *Carry*, is $2\times3^n + 2\times3^{n+1} = 8\times3^n$. However, the input range is limited to $3\times2\times3^n = 6\times3^n$. As a result, *4:2* compressors, which do not waste their output range, are inevitably required to design high-order ternary compressors. For example, it is unlikely to design an efficient ternary *8:3* compressor by means of merely half and full adders. Without *4:2* compressors, redundant outputs will be generated.

# IV. Conclusion

The studies in this paper show that a complete TFA is not required in the arithmetic blocks of ternary subtractor and multiplier. They can be designed without the need of a complete TFA. Compressor is another important arithmetic block in which full adders are widely used. Some high-order ternary compressors have been presented in this paper as well. Although there are some TFAs in the proposed designs, none of the compressors depends on the operation of a complete TFA due to the fact that several PPs in a ternary multiplier belong to the number set of $\{0, 1\}_3$. In this situation, partial TFAs will suffice to design wide ternary compressors.

Since a complete TFA is no longer needed (at least in the aforementioned applications), circuit designers can focus on designing partial TFAs such the ones in [9, 19]. The truth table of a partial TFA has nine fewer rows than a complete one. This will certainly lead to elimination of some transistors. Consequently, TFAs can have higher performance, more compactness, lower power consumption, and higher speed. This is a very consequential result in VLSI design.